\documentclass [11pt]{article} 
\usepackage{amsmath,amsthm,amsfonts,amscd,eucal,latexsym,amssymb} 
\usepackage{epsfig}
\def\sp{\hskip -5pt} 
\def\spa{\hskip -3pt}

\newsymbol\bt 1202           
\def\bC{{\mathbb C}}           

\def\bR{{\mathbb R}} 
\def\bS{{\mathbb S}}

\def\bZ{{\mathbb Z}} 
 
\newsymbol\rest 1316         
 
\def\beq{\begin{eqnarray}}
\def\eeq{\end{eqnarray}}
\def\pa{\partial}
\def\at{\left(}               
\def\ct{\right)}              
\newcommand{\ca}[1]{{\cal #1}}         

\def\ka{\kappa}

\def\Om{\Omega}

\begin{document} 
 
\hfill{\sl UTF 452/UTM 641} 
\par 
\bigskip 
\par 
\rm

 
\par 
\bigskip 
\LARGE 
\noindent 
{\bf QFT holography near the horizon of Schwarzschild-like spacetimes.} 
\bigskip 
\par 
\rm 
\normalsize 
 
 
 
\large 
\noindent {\bf Valter Moretti$^{1,2}$ and Nicola Pinamonti$^{3}$}

\noindent 
$^1$ Department of Mathematics, E-mail: moretti@science.unitn.it\\ 
$^2$ I.N.F.N. Gruppo Collegato di Trento\\ 
$^3$ Department of Physics, E-mail: pinamont@science.unitn.it\\ 
\rm\large \large 
 
\noindent 
 University of Trento, 
 Faculty of Science,
via Sommarive 14,
I-38050 Povo (TN),
Italy. 
\large 
\smallskip

\rm\normalsize 
 

 
\par 
\bigskip 
\par 
\hfill{\sl April 2003} 
\par 
\medskip 
\par\rm

\noindent
\small 
{\bf Abstract:}
\noindent It is argued that free QFT can be defined on the event horizon of a Schwarzschild-like
spacetime and that that theory is unitarily and algebraically equivalent to QFT in the bulk
(near the horizon). Under that unitary equivalence the bulk hidden $SL(2,\bR)$ symmetry 
found in a previous work becomes manifest on the event horizon, it being induced by a group
of horizon diffeomorphisms. The class of generators of that group can be enlarged to include
a full Virasoro algebra of fields defined on the event horizon. These generators have
a quantum representation in QFT on the event horizon and thus in the bulk.   
\normalsize
\bigskip

\noindent {\bf 1}. {\em Introduction}. 
A number of papers has been concerned with the issue of the statistical origin of black-hole entropy. 
Holographic principle \cite{thoo93, thoo95, suss95} arose by the idea  
that gravity near the horizon should be described by a low dimensional theory with a higher dimensional group
of symmetry. The correspondence between quantum field theories of different dimensions 
was conjectured by Maldacena \cite{mald98} using the machinery of 
string theory: There is a correspondence between quantum field theory in
a, asymptotically $AdS$,  $d+1$ dimensional spacetime  (the ``bulk'')
 and a conformal theory in a $d$
dimensional manifold (the (conformal) ``boundary'' at spacelike infinity).  
Witten \cite{witt98} described the above correspondence  in terms of relations
of observables of the two theories.
Rehren proved rigorously some holographic results for free quantum fields in a $AdS$ 
background, establishing a correspondence between bulk observables and boundary observables 
without employing string machinery \cite{rehr00a,rehr00b}. 
To explain the correspondence one should notice that the 
conformal group which 
acts in the $d$-dimensional  $AdS_{d+1}$ boundary can be
 realized as the group of the isometries of the $AdS_{d+1}$ bulk.
In this way,  the bulk-boundary correspondence has a geometric nature. 
The boundary of a Schwarzschild spacetime  (dropping the boundary at
infinity) is the event horizon of the black hole. The AdS correspondence has been used
directly in Minkowski spacetime for massless particle in \cite{saso01} with the help
of the optical metric.
Is there any bulk-boundary correspondence in a manifold containing a Schwarzschild-like black hole?
Two-dimensional Rindler spacetime embedded in Minkowski spacetime approximates the nontrivial part
of the spacetime structure  near a bifurcate horizon as that of a Schwarzschild black hole embedded in Kruskal spacetime.
In that context,  we have argued in a recent work \cite{mopi02} that free quantum field theory  in
two-dimensional Rindler space  presents a ``hidden'' $SL(2,\bR)$ symmetry: The theory turns out
to be invariant under a unitary representation of $SL(2,\bR)$ but such a quantum symmetry cannot be induced
by the geometric background. 
$SL(2,\bR)$ is the group of symmetry of the zero-dimensional conformal field theory in the sense 
of \cite{DFF}, so,  as
for the case of AdS spacetime, it suggests the existence of a possible correspondence between quantum field
theory in Rindler space  and a conformal field theory defined on its event horizon.
In this  letter we illustrate the basic results that can be found in a forthcoming technical 
paper \cite{mopi03} where we have shown that  it is possible to build up the wanted 
correspondence of a free quantum theory defined in the bulk and a quantum field theory 
defined on the event horizon of a two dimensional Rindler space. Other involved results are that
the $SL(2,\bR)$ symmetry reveals a clear geometric meaning if it is examined on the horizon
and, in that context, a whole Virasoro algebra of symmetries arises.\\
Some overlap with our results is present in the literature. Guido, Longo, Roberts and Verch
\cite{GLRV01} discussed in some detail the extent to which an algebraic QFT on a spacetime with a bifurcate
Killing horizon induces a conformal QFT on that bifurcate Killing horizon.
Along a similar theme, Schroer and Wiesbrock \cite{SW00} have studied the relationship between QFTs on
horizons and QFTs on the ambient spacetime. They even use the term 
``hidden symmetry'' a sense similar as we do here and we done in \cite{mopi02}. In related follow-up works by
Schroer \cite{S01} and by Schroer and Fassarella \cite{SF01} the relation to holography 
and diffeomorphism covariance is also discussed.\\

\noindent {\bf 2}. {\em Hidden $SL(2,\bR)$ symmetry}.  Consider a general Schwarzschild-like metric (namely a static black hole metric
with bifurcate event horizon), 
 $ds^2_{\bf S} = -A(r)dt^2 +{A^{-1}(r)} {dr^2} + r^2 d\Om^2$, $\Om$ denoting angular coordinates. 
Near the horizon ($r=r_h$), the  nonangular part of the metric 
 reduces to the metric of a two-dimensional Rindler wedge ${\bf R}$,
$ds^2_{\bf R} = -{\kappa}^2 y^2 dt^2 + dy^2$ 
with $A'(r_h)=2\ka$, and $\ka y^2=4(r-r_h)$.
Also dropping the angular coordinates,
let us consider a free Klein-Gordon scalar field $\phi$ with motion equation
$-\partial^2_t \phi  + {\kappa}^2 \left(y \partial_yy \partial_y  -  
 y^2 m^2 \right) \phi =0$. To built up the one-particle Hilbert space referred to the quantization with respect to 
the Rindler Killing time $t$,
 any real solution $\psi$ of the K-G equation must be decomposed in $\partial_t$-stationary modes as follows
\begin{eqnarray}
\psi(t,y) = \int_{0}^{+\infty} \: \sum_\alpha\Phi^{(\alpha)}_{E}(t,y) \tilde\psi^{(\alpha)}_+(E)\: dE + c.c. \label{wave}
\end{eqnarray}
$E\in [0,+\infty) = \bR^+$ is an element of the spectrum of the Rindler Hamiltonian $H$ associated with $\partial_t$
evolution.
Concerning the index $\alpha$ we distinguish between two cases: if $m>0$ there is a unique mode 
$\Phi^{(\alpha)}_{E}=\Phi_{E}$ whose expression is
$\sqrt{2E \sinh(\pi E/\kappa)}/\sqrt{2\pi^2\kappa\, E}\:{e^{-iEt}}\:
K_{iE/\kappa}(my)$.
If $m=0$ there are two values of $\alpha$, corresponding to 
$in$going  and $out$going modes, $\Phi^{(in)/(out)}_E$ whose expression are $e^{-iE(t\pm \ln{(\ka
y)/\ka})}/\sqrt{4\pi E}$.
If $m>0$ there is no energy degeneration and the one-particle Hilbert space $\ca{H}$ generated by the 
positive frequency part of the decomposition above is isomorphic to $L^2(\bR^+,dE)$.
In the other case ($m=0$), twofold degeneracy implies that $\ca{H} \cong L^2(\bR^+,dE)\oplus L^2(\bR^+,dE)$. Quantum field operators, acting  in the symmetrized Fock space 
$\ca{F}(\ca{H})$ and referred to the Rindler vacuum $|0\rangle$ -- that is $|0\rangle_{in}\otimes |0\rangle_{out}$ 
if $m=0$ -- read 
\beq
\hat{\phi}(t,y)=\int_0^\infty\sum_\alpha
\Phi^{(\alpha)}_E(t,y)a_{E\alpha}+\overline{\Phi^{(\alpha)}_E(t,y)}a_{E\alpha}^\dagger dE.
\eeq
As usual, the causal propagator $\Delta$ satisfies
$[\hat{\phi}(x),\hat{\phi}(x')]=-i\Delta(x,x')$.\\
In \cite{mopi02} we have found that, if $m>0$, $\ca{H}$ is irreducible under a unitary 
representation  of ${SL}(2,\bR)$ generated by (self-adjoint extensions) of the operators 
$iH, iD, iC$ (which enjoy the commutation relations of the Lie algebra of $SL(2,\bR)$), with
\begin{equation}
H := E \:,\:\:\:\:\:
D := -i\left(\frac{1}{2} + E\frac{d \:}{d E}\right)\:, \:\:\:\:\:
C := -\frac{d \:}{d E} E\frac{d \:}{d E} +
\frac{(k-\frac{1}{2})^2}{E}\:.\label{ge3F}  
\end{equation} 
$k$ can arbitrarily be fixed in $\{1/2,1,3/2,\ldots\}$.
 See \cite{mopi03} for details on domains an all that. If $m=0$ 
and so $\ca{H} \cong L^2(\bR^+,dE)\oplus L^2(\bR^+,dE)$,
an analogue representation exists in each space $L^2(\bR^+,dE)$. Making use of  Heisenberg representation
it is simply proven that the algebra generated by $H,D,C$, with depending-on-time coefficients, is made of
constant of motions \cite{mopi02,mopi03,DFF}. Thus $SL(2,\bR)$ is a symmetry of the one-particle system 
(that can straightforwardly be extended to the free 
quantum field in Fock space). The crucial point is that the found symmetry is {\em hidden}: it cannot be induced by the background
geometry since the Killing fields of Rindler spacetime enjoy a different Lie algebra from that of $H,D,C$.\\

\noindent {\bf 3}. {\em Fields on the horizon}. Let us  to investigate the nature of the found  
symmetry exactly on the event horizon
assuming ${\bf R}$ to be naturally embedded in a Minkowski spacetime. In particular we want to investigate
its geometrical nature, if any, on the event horizon.
\begin{center}
\epsfig{file=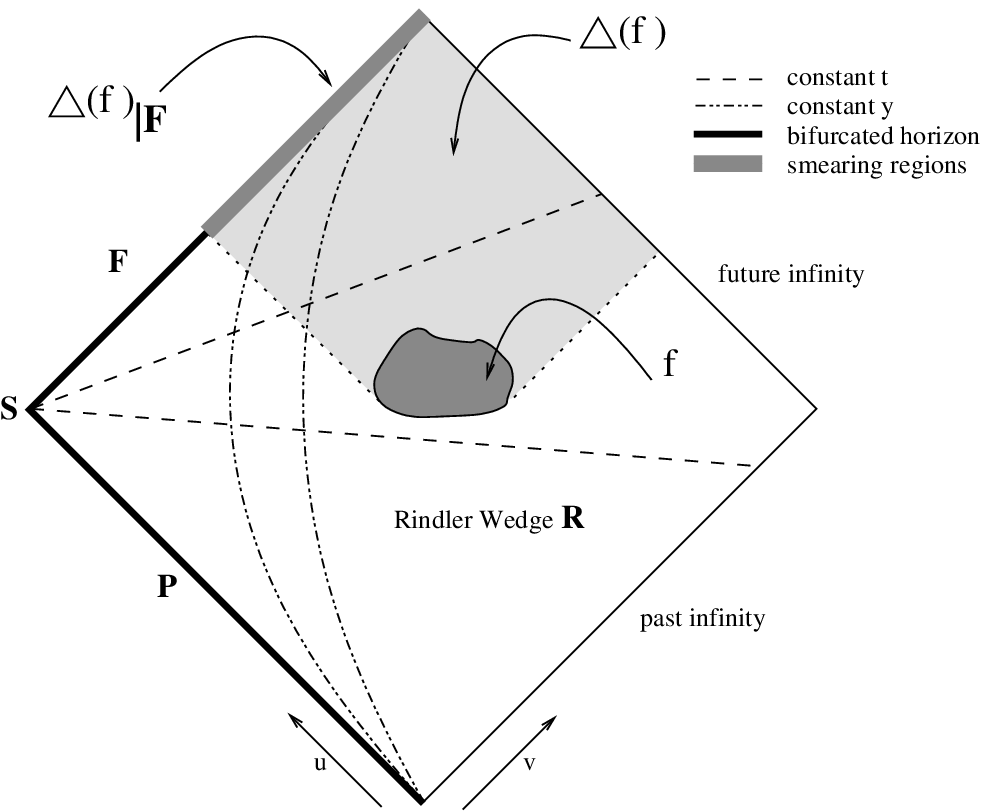,width=.6\textwidth}
\end{center}
 (Rindler) {\em light coordinates} 
$u= t-{\log(\ka y)}/{\ka}$, $v = t+{\log(\ka y)}/{\ka}$ (where $u,v\in \bR$) 
cover the (open) Rindler space  ${\bf R}$. Separately, $v$ is
well defined on the future horizon ${\bf F}$, $u\to +\infty$, and $u$ is well defined  on the past horizon
${\bf P}$, $v\to -\infty$ (see figure). Take the wavefunction in (\ref{wave}) and consider the limit on
the future horizon $u\to +\infty$. That is equivalent to restrict the wavefunction on the event horizon when
it is considered as a wavefunction in Minkowski spacetime, obtaining
\begin{eqnarray}
\psi(v)  =  \int \frac{e^{-iEv}}{\sqrt{4\pi E}}
e^{i\rho_{m,\ka}(E)}\tilde\psi_+(E)\: dE + \mbox{c.c.} \label{limitH+}
\end{eqnarray}
$e^{i\rho_{m,\ka}(E)}$ is a pure phase (see \cite{mopi03} for details).
In coordinate $u\in \bR$, the restriction of $\psi$ to ${\bf P}$ is similar  with the $v$
replaced for $u$ and  $\rho_{m,\ka}(E)$ replaced by $-\rho_{m,\ka}(E)$. 
If $m=0$ the restrictions to ${\bf F}$ and ${\bf P}$ read respectively
\beq
\psi(v)  =  \int \frac{e^{-iEv}}{\sqrt{4\pi E}}
\tilde\psi^{(in)}_+(E)\: dE + \mbox{c.c.} \:\:, \qquad 
\psi(u)  =\int \frac{e^{-iEu}}{\sqrt{4\pi E}} \tilde\psi^{(out)}_+(E)\: dE + \mbox{c.c.} \:\:
\eeq
Discarding the phase it is possible to consider the following real
``field on the future Horizon'':
\begin{eqnarray}
\varphi(v)  =  \int_{\bR^+} \frac{e^{-iEv}}{\sqrt{4\pi E}}
 \tilde\varphi_+(E)\: dE + \int_{\bR^+} \frac{e^{+iEv}}{\sqrt{4\pi E}}
 \overline{\tilde\psi_+(E)}\: dE \label{fieldonhorizon}
\end{eqnarray}
as the basic object in defining a quantum field theory on the future event horizon. The same can be done 
for the past event horizon. The one-particle Hilbert space ${\cal H}_{\bf F}$ is defined 
as the space generated by positive frequency parts $\tilde\psi_+(E)$ and turns out to be 
isomorphic to $L^2(\bR^+,dE)$ once again. The field operator reads, on the symmetrized Fock space ${\cal F}({\cal H}_{\bf F})$
with vacuum $|0\rangle_{\bf F}$, 
\beq
\hat{\phi}_{\bf F}(v)=\int_0^\infty \frac{e^{-iEv}}{\sqrt{4\pi E}} a_{E}+ \frac{e^{iEv}}{\sqrt{4\pi E}} a_{E}^\dagger dE\:.
\eeq
The causal propagator $\Delta_{\bf F}$ is defined by imposing $[\hat{\phi}(v),\hat{\phi}(v')]=-i\Delta_{\bf F}(v,v')$ and  it takes the form  $(1/4)sign(v-v')$. In spite of the 
absence of any 
motion equation the essential features of free quantum field theory are preserved by that definition as proven in \cite{mopi03}.
 Degeneracy of the metric on the horizon prevents from smearing  field operators by functions due to the 
 ill-definiteness of the induced volume measure. However, employing the symplectic approach \cite{Wald}, 
 a well-defined smearing-procedure is that of field operators and exact $1$-forms $\eta=df$ where $f=f(v)$ vanishes
fast as $v\mapsto \pm\infty$. The integration of forms does not need any measure.
 In other words for a real exact $1$-form $\eta$ as said above
\beq
\hat{\phi}_{\bf F}(\eta)=\int_0^\infty \frac{dE}{\sqrt{4\pi E}}
\left(\int_\bR e^{-iEv}\eta(v)\right) a_{E}+ 
\left(\int_\bR e^{iEv} \eta(v)\right) a_{E}^\dagger 
\eeq
is well defined and diffeomorphism invariant.
 In a suitable domain the map $\eta(v) \mapsto\Delta_{\bf F}(\eta) = \frac{1}{4}\int_\bR sign(v-v')\eta(v') = \psi_\eta(v)$
 defines a one-to-one correspondence between exact one-forms and horizon wavefunctions of the form (\ref{wave})
 and $\eta = 2d\psi_\eta $.
 Finally, similarly to usual quantum field theory \cite{Wald},  it holds  
 $$[\hat{\phi}_{\bf F}(\eta), \hat{\phi}_{\bf F}(\eta')] = -i \Delta_{\bf F}(\eta,\eta') = \int_{\bf F} \psi_{\eta'}d\psi_{\eta}- 
  \psi_{\eta} d\psi_{\eta'}.$$
 The last term define a diffeomorphism-invariant symplectic form on horizon vavefunctions.\\
 
 \noindent {\bf 4}. {\em Unitary and algebraic holographic theorems}.
It is possible to prove the existence of a unitary equivalence between the theory in the bulk and that on the horizon in the sense we
are going to describe. Consider the case $m>0$ and the future horizon ${\bf F}$. 

{\em {\bf Theorem 1}. There is a unitary map
 $U_{\bf F}: {\cal F}({\cal H}) \to {\cal F}({\cal H}_{\bf F})$ such that $U_{\bf F} |0\rangle = |0\rangle_{\bf F}$
 and $U^{-1}_{\bf F} \hat\phi_{\bf F}(\eta) U_{\bf F} = \hat\phi(f)$ for any smooth compactly supported function $f$
 used to smear the bulk field, $\eta = 2d(\Delta(f)\spa\rest_{\bf F})$.}  (See figure.)\\
 Details on the construction of $U_{\bf F}$ are supplied in \cite{mopi03}, here we give only the main idea.
 Take $f$ as said and consider the associated bulk wavefunction $\psi_f=\Delta(f)$, restrict $\psi_f$ to ${\bf F}$
 obtaining a horizon vavefunction as in (\ref{limitH+}) with positive frequency part $e^{i\rho_{m,\ka}}(E) \tilde{\psi}_{f+}(E)$.
 Then define a horizon vavefunction $\varphi_f$ as in (\ref{fieldonhorizon}) with $\tilde\varphi_+$ replaced by 
 $\tilde{\psi}_{f+}$. 
 It is clear that the  map $\psi_f \mapsto \varphi_f$ corresponds to a unitary operator from ${\cal H}$ to ${\cal H}_{\bf F}$. 
 That is, by definition $U_{\bf F}\sp\rest_{\cal H}$. Imposing $U_{\bf F} |0\rangle = |0\rangle_{\bf F}$, by taking
 tensor products of $U_{\bf F}\sp\rest_{\cal H}$, this map extends to a unitary map 
 $U_{\bf F}: {\cal F}({\cal H}) \to {\cal F}({\cal H}_{\bf F})$.
 Finally, by direct inspection one finds that, if  $\eta= 2d \varphi_f$, one also has
 $U^{-1}_{\bf F} \hat\phi_{\bf F}(\eta) U_{\bf F} = \hat\phi(f)$.\\
 The same procedure can be used to define an analogous unitary operator referred to ${\bf P}$.
 If $m=0$ two unitary operators arises.  One is $V_{\bf F}: {\cal F}({\cal H}_{in})  \to {\cal F}({\cal H}_{\bf F})$ 
 such that $V_{\bf F} |0\rangle_{in} = |0\rangle_{\bf F}$
 and $V^{-1}_{\bf F} \hat\phi_{\bf F}(\eta_f) V_{\bf F} = \hat\phi_{in}(f)$. ${\cal H}_{in}$ is the bulk Hilbert space
 associated with the ingoing modes ans $\hat\phi_{in}(f)$ is the part of bulk field operator built up using only 
 ingoing modes. The other unitary operator  $V_{\bf P}: {\cal F}({\cal H}_{out})  \to {\cal F}({\cal H}_{\bf P})$
 plays an analogous r\^ole with $in$ replaced for $out$ and ${\bf F}$ replaced for ${\bf F}$ everywhere.
 (More generally $V_{\bf P}\otimes V_{\bf F} : {\cal F}({\cal H})  \to {\cal F}({\cal H}_{\bf P})\otimes {\cal F}({\cal H}_{\bf F})$
 define a unitary operator which transforms the vacuum states into vacuum states and  field operators
 into field operators.)
 As a consequence of the cited theorem, e.g. if $m>0$, one has the invariance of vacuum expectation values:
 $$_{\bf F}\langle 0| \hat\phi_{\bf F}(\eta_{1})\cdots \hat\phi_{\bf F}(\eta_{n})|0\rangle_{\bf F}
 = \langle 0| \hat\phi(f_1)\cdots \hat\phi({f_n})|0\rangle\:.$$
 Similarly to the extent in the bulk case, one focuses on the algebra
${\cal A}_{\bf F}$  of linear combinations of product of field operators $\hat\phi_{\bf F}(\omega)$ varying $\omega$
in the space of allowed complex $1$-forms. We assume that ${\cal A}_{\bf F}$ also contain the unit operator $I$. 
The Hermitean elements of ${\cal A}_{\bf F}$ are the natural 
observables associated with the horizon field.
 From an abstract point of view the found algebra is a unital $*$-algebra of formal operators ${\phi}_{\bf F}(\eta)$
with the additional properties $[{\phi}_{\bf F}(\eta), {\phi}_{\bf F}(\eta)] = -i \Delta_{\bf F}(\eta,\eta')$,
${\phi}_{\bf F}(\eta)^*= {\phi}_{\bf F}(\overline{\eta})$ and linearity in the form $\eta$\footnote{The analogous algebra
of operators in the bulk fulfill the further  requirement $\phi(f)=0$ if (and only if ) $f=Kg$, $K$ being the Klein-Gordon operator.
No analogous requirement makes sense for ${\cal A}_{\bf F}$ since there is no equation of motion on the horizon.}.
${\cal A}_{\bf F}$ can be studied no matter any operator representation in any Fock space. Operator representations
are obtained via GNS theorem once an algebraic state has been fixed \cite{Wald}. In the case $m>0$
we get the following result which is independent from any choice of vacuum state and Fock representation.
The proof can be found in
\cite{mopi03}.
${\cal A}_{\bf R}$ denotes the unital $*$-algebra of associated with the bulk field operator.

  {\em {\bf Theorem 2}. There is a unique 
 injective unital $*$-algebras homomorphism $\chi_{\bf F}: {\cal A}_{\bf R} \to {\cal A}_{\bf F}$ such that 
 $\chi_{\bf F}(\phi(f)) = \phi_{\bf F}(\eta_f)$, where $\eta = 2d(\Delta(f)\spa\rest_{\bf F})$. Moreover in GNS representations in the respectively associated 
 Fock spaces ${\cal F}({\cal H})$, ${\cal F}({\cal H}_{\bf F})$ built up over $|0\rangle$ and $|0\rangle_{\bf F}$ 
 respectively, $\chi_{\bf F}$ has a unitary implementation and reduces to $U_{\bf F}$.}\\
Notice that, in particular $\chi_{\bf F}$ preserves the causal propagator, in the sense that it must be
$-i\Delta(f,g) = [\phi(f),\phi(g)]I = [\phi(f),\phi(g)]\chi_{\bf F}(I)= \chi_{\bf F}([\phi(f),\phi(g)]I)=
[\chi(\phi(f)),\chi(\phi(g))]\\ = [\phi_{\bf F}(f),\phi_{\bf F}(g)] = -i\Delta_{\bf F}(\eta_f,\eta_g)$.\\
Analogous algebraic homomorphism theorems can be given for ${\bf P}$ and the massless case \cite{mopi03}.\\

\noindent {\bf 5}. {\it The $SL(2,\bR)$ symmetry becomes manifest on the horizon.} Consider quantum field theory 
on ${\bf F}$, but the same result holds concerning ${\bf P}$. In ${\cal H}_{\bf F}\cong L^2(\bR^+,dE)$
define  operators 
$H_{\bf F},D_{\bf F},C_{\bf F}$ as the right-hand side of the equation that respectively defines $H,D,C$
in (\ref{ge3F}). 
Exactly as in the bulk case,  operators $iH_{\bf F},iD_{\bf F},iC_{\bf F}$ generate a unitary $SL(2,\bR)$ 
representation $\{{\cal U}_g\}_{g\in SL(2,\bR)}$. Hence, varying $g\in SL(2,\bR)$,
$U_g = (U_{\bf F}\spa\rest_{\cal H})^{-1} \:{\cal U}_g \:U_{\bf F}\spa\rest_{\cal H}$ define a representation of 
 $SL(2,\bR)$ for the system in the bulk. By construction 
 $(U_{\bf F}\spa\rest_{\cal H})^{-1} \:H_{\bf F} \: U_{\bf F}\spa\rest_{\cal H}= H$. As a consequence
 every $U_{g}$ turns out to be  a $SL(2,\bR)$ symmetry of the bulk system  and the group of these symmetries 
 is unitary equivalent to that generated by 
 $iH,iD,iC$. In particular the one-parameter group associated with $H_{\bf F}$ generates $v$-displacements
 of horizon wavefunctions which are equivalent, under the action of $U_{\bf F}$, to $t$-displacements
 of bulk wavefunctions. Now, it make sense to investigate the {\em geometrical nature} of the $SL(2,\bR)$
 representation $\{{\cal U}_g\}$ that, as we said, induces, up to unitary equivalences, 
 the original $SL(2,\bR)$ symmetry in the bulk. In fact it is possible to prove that:
 
 {\em {\bf Theorem 3}. If $k=1$ in (\ref{ge3F}), the action of every ${\cal U}_g$ on a state $\tilde \varphi_+=\tilde\varphi_+(E)$
 is essentially equivalent to the action of a corresponding 
 ${\bf F}$-diffeomorphism on the associated (by (\ref{fieldonhorizon})) horizon wavefunction $\varphi$.
 More precisely, take a matrix $g\in SL(2,\bR)$ and 
  $\varphi= \varphi(v)$ in a suitable space of horizon
 wavefunction (see \cite{mopi03}). Let $\tilde\varphi_+=\tilde\varphi_+(E)$ be the positive frequency part 
 of $\varphi$.
 The wavefunction $\varphi_g$ associated with ${\cal U}_g\tilde\varphi_+$ reads
\beq
\varphi_g(v)= \varphi\at \frac {av+b}{cv+d}\ct - \varphi\left(\frac{b}{d}\right) ,\qquad g^{-1}=\begin{pmatrix}a&b\\c&d
\end{pmatrix}. 
\eeq}
The term $- \varphi({b}/{d})$ assures that $\varphi_g$ vanishes as $v\to \pm\infty$. Notice 
that the added term disappears when referring to $d\varphi$ rather than $\varphi$.
The group of diffeomorphisms of ${\bf F}$, i.e. the real line\footnote{Actually one has to consider the projective line
${\bf F}\cup \{\infty\}$.}, 
\beq
v\mapsto \frac{av+b}{cv+d} ,\qquad g =\begin{pmatrix}a&b\\c&d
\end{pmatrix} \in SL(2,\bR)
\eeq
can be obtained by composition of one-parameter subgroups associated with the following three vector fields
on ${\bf F}$:
$\partial_v, v\partial_v, v^2\partial_v$. It is simply proven that the Lie brackets of those fields is
 a realization of the Lie algebra of $SL(2,\bR)$. Moreover, it turns out that \cite{mopi03}:

{\em {\bf Theorem 4}  
(a) If $k=1$ in (\ref{ge3F}), the unitary one-parameter group generated by $iH_{\bf F}$ 
 is associated, through Theorem 3, to the one-group of ${\bf F}$-diffeomorphisms 
 generated by $\partial_v$, (b) the unitary one-parameter group generated by $iD_{\bf F}$ is 
 associated to the one-parameter group of ${\bf F}$-diffeomorphisms 
 generated by $v\partial_v$ and (c) the unitary one-parameter group generated by $iC_{\bf F}$ 
 is associated to the one-group of ${\bf F}$-diffeomorphisms generated by $v^2\partial_v$.}\\

\noindent {\bf 6}. {\em Appearance of Virasoro algebra}. 
The bulk $SL(2,\bR)$-symmetry is manifest when examined on the event horizon, in the sense that 
it is induced by the geometry.  The Lie algebra generated by  vector fields $\partial_v, v\partial_v, v^2\partial_v$ 
play a crucial r\^ole in proving this fact. That algebra can be extended to include all the class of fields
defined on the event horizon $\{{\cal L}_n\}_{n\in \bZ}$ with ${\cal L}_n = v^{n+1}\pa_v$. It is interesting to notice that these fields
enjoy Virasoro commutation relations  without central charge, $\{{\cal L}_n ,{\cal L}_m\}=(n-m){\cal L}_{n+m}$.
A natural question arises:\\ {\em Is it possible to give a quantum representation of these generators in the sense of 
Theorem 4?}\\
 At least formally, the answer is positive. Indeed, by employing Theorem 4 one finds out that the infinitesimal
action of the one parameter group of diffeomorphisms generated by ${\cal L}_n$ on a horizon 
wavefunction $\varphi= \varphi(v)$ is equivalent to the action of an anti-Hermitean operator $L_n$ 
on the positive frequency part $\tilde\varphi_+=\tilde\varphi_+(E)$. $L_n$ is defined as, respectively for $n\geq-1$
and $n<-1$,
\begin{eqnarray}
(L_n \tilde\varphi_{+})(E) &:=& i^{n+2} \sqrt{E}\frac{d^{n+1}}{dE^{n+1}}\sqrt{E}\tilde\varphi_{+}(E) \label{ngeq-1}\:,\\
(L_n \tilde\varphi_{+})(E) &:=&  -i^{-(n+2)} \sqrt{E}\int_0^E \sp\sp dE_1 \int_0^{E_1} \sp\sp dE_2 \:\cdots 
\int_0^{E_{-(n+2)}} \sp \sp\sp \sp\sp \sp\sp dE_{-(n+1)}\sqrt{E_{-(n+1)}}\tilde\varphi_{+}(E_{-(n+1)}) \label{n<-1}\:.
\end{eqnarray}
Those operators are at least anti-Hermitean on suitable domains and enjoy Virasoro commutation rules
$[{L}_n, {L}_m]= (n-m)\: {L}_{n+m}$.\\

\noindent {\bf 7}. {\em Four dimensional case}. 
Up to now we have investigated only the two dimensional spacetimes, but it is possible to extend our results
to a four dimensional case which better approximates the Schwarzschild extent. 
For this purpose consider the 
near-horizon approximation of a Schwarzschild-like spacetime without discarding the angular variables $\theta,\phi$,
so that 
$ds^2 = -{\kappa}^2 y^2 dt^2 + dy^2+ r_h^2 d\Om^2$. Every field takes an angular part described by the usual spherical harmonics
$Y^l_m(\theta, \phi)$. QFT in the bulk involves the one-particle Hilbert space 
$\oplus_{l=0}^{\infty} \bC^{2l+1}\otimes {\cal H}_l$ with ${\cal H}_l \cong L^2(\bR^+,dE)$ if $l>0$,
 $\bC^{2l+1}$ being the space at fixed 
total angular momentum $l$ and ${\cal H}_0 \cong L^2(\bR^+,dE)$ in the massive case 
but ${\cal H}_0 \cong L^2(\bR^+,dE)\oplus L^2(\bR^+,dE)$ in the massless case. For wavefunctions 
with components in a fixed space $\bC^{2l+1}\otimes L^2(\bR^+,dE)$ Klein-Gordon equation 
reduces to the two-dimensional one with a positive contribution to the mass depending on $l$. 
Quantum field theory can be constructed on the future horizon ${\bf F}\cong \bS^2\times \bR$.
The appropriate causal propagator reads $$\Delta_{\bf F}(x,x') = (1/4) sign(v-v')
\delta(\theta-\theta')\delta(\phi-\phi') \sqrt{g_{\:\bS^2}(\theta,\phi)}\:.$$ The horizon field operator $\hat\phi_{\bf F}$
 has to be smeared with $3$-forms as
$df(v,\theta,\phi) \wedge d\theta \wedge d\phi$ and Theorems 1 and 2, at least in the massive case,
 can be restated as they stand for the two-dimensional case. Theorems 3 and 4 hold true at fixed angular variables.\\

\noindent {\bf Acknoledgments}. The authors are grateful to R.Verch for kind and useful suggestions.

\end{document}